\documentclass[pre,twocolumn,aps]{revtex4-1}%
\usepackage{color}
\usepackage{graphicx}
\usepackage[english]{babel}
\usepackage{amsmath}
\usepackage{amsfonts}
\usepackage{amssymb}%
\usepackage{dsfont} 

\usepackage[dvipsnames]{xcolor}
\usepackage{color}

\newcommand{\be}{\begin{equation}}
\newcommand{\ee}{\end{equation}}
\newcommand{\bel}[1]{\begin{equation}\label{#1}}
\newcommand{\bea}{\begin{eqnarray}}
\newcommand{\eea}{\end{eqnarray}}
\newcommand{\balign}{\begin{align}}
\newcommand{\ealign}{\end{align}}
\newcommand{\ba}{\begin{array}}
\newcommand{\ea}{\end{array}}
\newcommand{\bfig}{\begin{figure}}
\newcommand{\efig}{\end{figure}}

\newcommand{\eref}[1]{(\ref{#1})}

\allowdisplaybreaks

\newcommand{\exval}[1]{\mbox{$\langle \, {#1}\, \rangle$}}

\newcommand{\floor}[1]{\lfloor{#1}\rfloor}

\newcommand{\rmd}{\mathrm{d}}
\newcommand{\rme}{\mathrm{e}}

\newcommand{\tr}{\mathop{\mathrm{tr}}\nolimits}

\newcommand{\comm}[2]{\mbox{$\left[{#1},\,{#2}\right]$}}

\newcommand{\R}{{\mathbb R}}

\newcommand{\Z}{{\mathbb Z}}

\begin{document}

\title{Charge-current correlation equalities for quantum systems far from equilibrium}
\author{D. Karevski$^1$, G.M. Sch\"utz$^{1,2}$}
\affiliation{(1) Laboratoire de Physique et Chimie Th\'eoriques,
Universit\'e de Lorraine, UMR CNRS 7019, B.P. 70239, F-54506 Vandoeuvre les Nancy
Cedex, France}
\affiliation{(2) Institute of Complex Systems II, Theoretical Soft Matter and Biophysics,
Forschungszentrum J\"ulich, 52425 J\"ulich, Germany}

\begin{abstract}
We prove that a recently derived correlation equality between conserved charges
and their associated conserved currents for quantum systems far from equilibrium 
[O.A. Castro-Alvaredo et al., Phys. Rev. X \textbf{6}, 041065 (2016)], is valid 
under more general conditions than assumed so far. Similar correlation identities, which 
in generalized Gibbs ensembles give rise to a current symmetry somewhat reminiscent of 
the Onsager relations, turn out to hold also in the absence of translation invariance, 
for lattice models, and in any space dimension, and to imply a symmetry of the 
non-equilibrium linear response functions.
\end{abstract}

\maketitle

\section{Introduction}

Of particular interest in the general context of transport far from thermal equilibrium
are the correlations between the conserved charges $Q^\alpha$ and their associated  
currents $J^\alpha_i$ in space direction $i$. We refer to the review by Spohn 
\cite{Spoh16} for a discussion from a broad perspective. Very recently, charge-current 
correlations in one-dimensional quantum integrable systems have been shown to play an 
important role in work on the Drude weight \cite{Doyo17,Uric18} and for generalized 
hydrodynamics \cite{Bert16,Cast16}.

Specifically, for the one-dimensional quantum case the global charge-current symmetry
\bel{CEglob}
\exval{Q^\alpha J^\beta}^c = \exval{J^\alpha Q^\beta}^c
\ee
for the connected correlation functions has been derived in \cite{Cast16} under quite 
general circumstances, viz., assuming only translation invariance of the stationary density 
matrix and the quantum Hamiltonian, a generic assumption on the 
decay of correlations, and, more significantly, commutativity of the stationary density 
matrix with the charges $Q^\alpha$. 

This result was subsequently generalized to a stronger local version 
$\exval{q^\alpha(x,t) j^\beta(0,0)}^c = \exval{j^\alpha(x,t) q^\beta(0,0)}^c$ \cite{xxx} 
which does {\it not} require the assumption of commutativity of the charges and which 
is valid for any decay of correlations with distance. The main aim 
of the present work is to derive related global and local charge-current correlation
equalities and to clarify the necessary and sufficient conditions under which such 
correlation equalities, including \eref{CEglob} and its local version, are valid. 

We start out from generic stationary one-dimensional lattice quantum systems of 
finite size with local conservation laws, without requiring translation invariance,
as detailed in Sec.~\ref{Sec:set}. The main results are derived in 
Sec.~\ref{Sec:ccc}, first very generally and then more specifically
under various additional generic conditions imposed on the physical system. 
Some simple consequences for symmetries of far-from-equilibrium linear response functions 
are indicated in Sec.~\ref{Sec:LR}.

All results derived below are straightforwardly extended to higher dimensions
by projection on one space coordinate and going through the same steps as below for 
each space coordinate. Furthermore, the results are valid in analogous form also for 
dissipative quantum systems where the time evolution of the density matrix is generated 
by a Lindblad quantum master equation, and for purely classical stochastic systems 
with Markovian dynamics for the probability distribution. However,
to avoid heavy notation and to expose clearly the essential ingredients that lead to
charge-current correlation equalities, we stick to the one-dimensional quantum context.

\section{The setting}
\label{Sec:set}

We consider a stationary many-body quantum system on a one-dimensional lattice 
of $L$ sites with Hamiltonian $H$. We shall not from the outset assume translation invariance,
but allow for non-translation-invariant stationary density matrices $\rho$ and/or spatially 
inhomogeneous dynamics encoded in $H$. The system is not assumed to be in thermal equilibrium.
Stationarity only means that we take expectations w.r.t. a density matrix $\rho$ that satisfies 
\bel{S1}
\mbox{\bf S1:} \quad\quad  \hspace*{2cm} \comm{\rho}{H} = 0. \hspace*{2cm}
\ee 

For observables $O$ we recall the definition
\bel{Heisenberg}
O(t) = \rme^{\frac{i}{\hbar} H t} O \rme^{-\frac{i}{\hbar} H t} 
\ee
of time-dependent operators in the Heisenberg picture.
We denote stationary expectation values and connected correlation functions
by $\exval{O}_L := \tr( \rho O)$ and $\exval{O_1(t) O_2}^c_L := 
\tr( \rho O_1(t) O_2) - \exval{O_1}\exval{O_2}$ resp., with the size-dependence 
indicated by the subscript $L$. 

Specifically, we consider a family of $n$ locally conserved 
charges, i.e., operators $q^\alpha_k$ 
that satisfy for $k\in\{1,\dots,L\}$ the discrete continuity equation
\bel{S2}
\mbox{\bf S2:} \quad \quad \hspace*{3mm} 
\frac{i}{\hbar} \comm{H}{q^\alpha_k(t)} = 
j^\alpha_{k-1}(t) - j^\alpha_{k}(t) \hspace*{8mm}
\ee
with the conserved currents $j_k^\alpha$ and the definition $j_0^\alpha:=j_L^\alpha$. 
Then the operators
\bel{conscharge} 
Q^\alpha = \sum_{k} q^\alpha_k
\ee
form a set of $n$ conserved charges $Q^\alpha = Q^\alpha(t)$. We remark
that for a non-translation invariant $H$ the operator $j^\alpha_{k-1}$ may not be
the lattice translation of $j^\alpha_{k}$. 
Nevertheless, the discrete continuity equation \eref{S2} alone implies
that the stationary current, denoted by $j^\alpha$, does not depend on $k$.

It is tacitly 
assumed that the charge and current operators are bounded so that all stationary
expectations of the charges $q^\alpha_k(t)$ and currents $j^\alpha_l(t)$
and all stationary correlations between them are finite for all 
system sizes $L$ and have well-defined thermodynamic limits.


In $d>1$ dimensions the lattice continuity equation 
for the locally conserved charges $q^\alpha_{\mathbf{k}}(t)$ at the lattice point $\mathbf{k} = (k_1,\dots,k_d)$ reads 
\be 
\frac{i}{\hbar} \comm{H}{q^\alpha_{\mathbf{k}}(t)} = 
\sum_{i=1}^d \left[j^{i,\alpha}_{\mathbf{k}^-_i}(t) - j^{i,\alpha}_{\mathbf{k}}(t)\right] 
\ee
with the conserved currents $j^{i,\alpha}_{\mathbf{k}}(t)$ in space direction $i$
and the shifted $i^{th}$ coordinate $\mathbf{k}^-_i := (k_1,\dots,k_i-1,\dots,k_d)$.
One considers the projected operators 
\bea 
q^\alpha_{k_i}(t) & = & \sum_{\mathbf{k}\setminus k_i} q^\alpha_{\mathbf{k}}(t) \\
j^{i,\alpha}_{k_i}(t) & = & \sum_{\mathbf{k}\setminus k_i} j^{i,\alpha}_{\mathbf{k}}(t) 
\eea
where the summations exclude the space coordinate $i$ and goes through the same calculations
as below for the one-dimensional case.

\section{Charge-current correlation equalities}
\label{Sec:ccc}

Specifically, we consider the time-dependent 
stationary correlation functions
\bea 
S^{\alpha\beta}_L(k,l,t) & := & \exval{q^\alpha_k(t) q^\beta_l(0)}^c_L, 
\label{Sab} \\
C^{\alpha\beta}_L(k,l,t) & := & \exval{j^\alpha_k(t) q^\beta_l(0)}^c_L,
\label{Cab} \\
\tilde{C}^{\alpha\beta}_L(k,l,t) & := & \exval{q^\alpha_k(t) j^\beta_l(0)}^c_L .
\label{tCab}
\eea
By identifying all lattice sites $k$ modulo $L$, the correlation functions can be defined 
for all $k,l\in\Z$ with periodicity $L$ for both space arguments $k,l$. 

\subsection{Results of general validity}

In this subsection we study relations between the charge-current correlation functions 
\eref{Cab} and \eref{tCab} 
that arise alone from S1 and S2, i.e., stationarity of the density matrix \eref{S1} and the 
conservation law \eref{S2}, without requiring translation invariance or any other specific 
property of $\rho$ or $H$. 

(i) With the Heisenberg representation \eref{Heisenberg} and the cyclic invariance of the trace,
one gets for the 
time derivative of the charge-charge correlation function \eref{Sab} the two expressions
\bea 
\dot{S}^{\alpha\beta}_L(k,l,t)
& = & \exval{(j^\alpha_{k-1}(t) - j^\alpha_{k}(t))q^\beta_l(0)}^c_L  
\label{jq1} \\
& = & - \exval{q^\alpha_{k}(t)(j_{l-1}^\beta(0) - j_{l}^\beta(0))}^c_L 
\label{jq2} 
\eea
from which one deduces by subtraction the fundamental charge-current
correlation equality
\bea
0 & = & C^{\alpha\beta}_L(k-1,l,t) - C^{\alpha\beta}_L(k,l,t) \nonumber \\
& & + \tilde{C}^{\alpha\beta}_L(k,l-1,t) - \tilde{C}^{\alpha\beta}_L(k,l,t).
\label{ccce1}
\eea
which is local in both coordinates $k$ and $l$ and which is the basis for further
considerations.

(ii) To explore consequences of this relation we consider the correlations involving 
the total charges $Q^\alpha$, viz.,
\bea 
A^{\alpha\beta}_L(k,t) & := & \sum_l C^{\alpha\beta}_L(k,l,t) \\
\tilde{A}^{\alpha\beta}_L(l,t) & := & \sum_k \tilde{C}^{\alpha\beta}_L(k,l,t) .
\eea
Because of the global charge conservation \eref{conscharge}, both averages 
$A^{\alpha\beta}_L(k,t)$ and $\tilde{A}^{\alpha\beta}_L(l,t)$ are trivially independent of time.
The local relations \eref{jq1} and \eref{jq2} then imply that both functions are independent
also of the space coordinate. This yields without further computation the 
charge-current correlation equalities
\bea
A^{\alpha\beta}_L(k,t) & = & \exval{j^\alpha_{k}(t) Q^{\beta}}^c_L 
\, = \,  \exval{j^\alpha_{0}(0) Q^{\beta}}^c_L 
\, =: \, a^{\alpha\beta}_L 
\label{ccce2} \\
\tilde{A}^{\alpha\beta}_L(l,t) & = & \exval{Q^\alpha(t)  j_{l}^\beta(0)}^c_L 
\, = \,  \exval{Q^\alpha  j_{0}^\beta(0)}^c_L 
\, =: \,  \tilde{a}^{\alpha\beta}_L
\label{ccce2t}
\eea
with constants $a^{\alpha\beta}_L$, $\tilde{a}^{\alpha\beta}_L$ that depend neither on 
$k$ nor on $t$. 

(iii) Next we consider the space averages
\bea 
B^{\alpha\beta}_L(r,t) & := & \frac{1}{L} \sum_k C^{\alpha\beta}_L(k,k+r,t) \\
\tilde{B}^{\alpha\beta}_L(r,t) & := & \frac{1}{L} \sum_k \tilde{C}^{\alpha\beta}_L(k,k+r,t) .
\eea
For examining the relationship between $B^{\alpha\beta}_L(k,t)$ and 
$\tilde{B}^{\alpha\beta}_L(l,t)$ we define the auxiliary function
\bel{gab} 
G^{\alpha\beta}_L(k,l,t)  :=  \sum_{k'=1}^k \left[\tilde{C}^{\alpha\beta}_L(k',0,t) 
- \tilde{C}^{\alpha\beta}_L(k',l,t) \right] 
\ee
and its space average 
\bel{fab} 
g^{\alpha\beta}_L(r,t)  :=  \frac{1}{L} \sum_k G^{\alpha\beta}_L(k,k+r,t)
\ee 
which allow for expressing both $C^{\alpha\beta}_L(k,t)$ and 
$\tilde{C}^{\alpha\beta}_L(l,t)$ in terms of $G^{\alpha\beta}_L(k,l,t)$ and the 
space averages $B^{\alpha\beta}_L(k,t)$ and $\tilde{B}^{\alpha\beta}_L(l,t)$ in terms
of $g^{\alpha\beta}_L(r,t)$. The auxiliary function $G^{\alpha\beta}_L(k,l,t)$ satisfies
$G^{\alpha\beta}_L(k,0,t) = G^{\alpha\beta}_L(0,l,t) = 0$ and periodicity property
\be 
G^{\alpha\beta}_L(k + mL,l+nL,t)=G^{\alpha\beta}_L(k,l,t)
\ee 
that is inherited from the periodicity of the correlation functions. Similarly, one has 
$g^{\alpha\beta}_L(r + mL,t)=g^{\alpha\beta}_L(r,t)$.

One gets from the definition \eref{gab} and from the doubly local relation \eref{ccce1}
\bea 
C^{\alpha\beta}_L(k,l,t) & = & C^{\alpha\beta}_L(0,l,t) \nonumber \\
& & + G^{\alpha\beta}_L(k,l,t) - G^{\alpha\beta}_L(k,l-1,t) 
\label{ccce1b}\\
\tilde{C}^{\alpha\beta}_L(k,l,t)  & = &  \tilde{C}^{\alpha\beta}_L(k,0,t) \nonumber \\
& & + G^{\alpha\beta}_L(k-1,l,t) - G^{\alpha\beta}_L(k,l,t) .
\label{ccce1c}
\eea
By setting $l=k+r$ in \eref{ccce1b} and \eref{ccce1c} and summing over $k$ one finds from
the charge-current correlation equalities
\eref{ccce2} and \eref{ccce2t} 
\bea
B^{\alpha\beta}_L(r,t) & = & \frac{1}{L} a^{\alpha\beta}_L  + g^{\alpha\beta}_L(r,t) - g^{\alpha\beta}_L(r-1,t)  \\
\tilde{B}^{\alpha\beta}_L(r-1,t) & = & \frac{1}{L} \tilde{a}^{\alpha\beta}_L  + g^{\alpha\beta}_L(r,t) - g^{\alpha\beta}_L(r-1,t)  
\eea
in terms of the space average \eref{fab}.
Thus we arrive at the charge-current correlation equality
\bel{ccce3}
B^{\alpha\beta}_L(r+1,t) -  \tilde{B}^{\alpha\beta}_L(r,t) 
= \frac{1}{L} \alpha^{\alpha\beta}_L \quad \forall r,t
\ee
with the constant $\alpha^{\alpha\beta}_L := a^{\alpha\beta}_L - \tilde{a}^{\alpha\beta}_L$.

The constant $\alpha^{\alpha\beta}_L$ is given by \eref{ccce2} and \eref{ccce2t}
\bel{alpha}
\alpha^{\alpha\beta}_L 
= \exval{j^\alpha_{0}(0) Q^{\beta}}^c_L  - \exval{Q^\alpha  j_{0}^\beta(0)}^c_L  
\ee
in terms of the stationary charge-current correlations for the global charges.
Notice that the independence of $r$ and $t$ allows for expressing $\alpha^{\alpha\beta}_L$
also as a stationary long-distance correlation as
\bel{alpha2} 
\alpha^{\alpha\beta}_L = L[B^{\alpha\beta}_L(\floor{L/2}+1,0) -  \tilde{B}^{\alpha\beta}_L(\floor{L/2},0)] 
\ee
where $\floor{x}\in\Z$ is the integer part of $x\in\R$.
This is a finite-size term that is generically small, but can be relevant for 
long-range interactions or non-local 
conserved charges. Also in the presence of stationary long-range
correlations at or below a quantum critical point the correlation may not be 
negligible.

As an aside we note without further comment that by \eref{jq1} and \eref{jq2} 
the auxiliary function $G^{\alpha\beta}(k,l,t)$ is related to the
structure function as
\bea
\dot{S}^{\alpha\beta}_L(k,l,t) 
& = & G^{\alpha\beta}_L(k-1,l,t)  + G^{\alpha\beta}_L(k,l-1,t) \nonumber \\
& & - G^{\alpha\beta}_L(k-1,l-1,t) - G^{\alpha\beta}_L(k,l,t).
\label{evo3} 
\eea
For the space average 
\be 
s^{\alpha\beta}_L(r,t) := \frac{1}{L}  \sum_k 
S^{\alpha\beta}_L(k,k+r,t)
\ee
one gets the evolution equation
\be 
\dot{s}^{\alpha\beta}_L(r,t) 
= g^{\alpha\beta}_L(r+1,t)  + g^{\alpha\beta}_L(r-1,t) - 2 g^{\alpha\beta}_L(r,t) .
\label{evo4} 
\ee

\subsection{Specializations}

The results \eref{ccce1}, \eref{ccce2}, \eref{ccce2t}, and \eref{ccce3} - \eref{alpha2} are valid without 
any conditions on the density matrix $\rho$ and on the Hamiltonian $H$, except that
all correlations are assumed to be bounded. Now we consider
some conditions of a general character and explore their consequences. 

\subsubsection{Decay of correlations}

We make the generic assumption of decay 
of correlations in the thermodynamic limit $L\to\infty$, i.e., for all $r,t$ we postulate
\be 
\label{C1}
\mbox{\bf C1:} \hspace*{5mm}
\lim_{r\to\infty} B^{\alpha\beta}_\infty(r,t) = \lim_{r\to\infty} \tilde{B}^{\alpha\beta}_\infty(r,t) = 0.
\ee 
This assumption is justified by the finite Lieb-Robinson
speed in non-relativistic quantum mechanics \cite{Lieb72}.

Decay of correlations implies $\alpha^{\alpha\beta}_L/L\to 0$ for $L\to\infty$ and 
therefore \eref{ccce3} yields the asymptotic charge-current correlation equality
\bel{ccce32}
B^{\alpha\beta}_\infty(r+1,t) = \tilde{B}^{\alpha\beta}_\infty(r,t) 
\ee
for the space averaged correlation function.

Under the slightly stronger condition
\bel{C1s} 
\mbox{\bf C1':} 
\lim_{L\to\infty} L[B^{\alpha\beta}_L(\floor{L/2}+1,0) 
-  \tilde{B}^{\alpha\beta}_L(\floor{L/2},0)] = 0 
\ee
on the decay of correlations 
one has $\alpha^{\alpha\beta}_L\to 0$ for $L\to\infty$. Then \eref{alpha} yields
\bel{ccce33} 
\exval{j^\alpha_{0}(0) Q^{\beta}}^c_\infty  = \exval{Q^\alpha  j_{0}^\beta(0)}^c_\infty.
\ee

We stress that no translation invariance is used to prove \eref{ccce32} and \eref{ccce33}.

\subsubsection{Translation invariance}

Now we consider the case where both $\rho$ and $H$ are translation invariant, i.e.,
for the lattice translation operator $T$ that transforms observables 
indexed by site $k$ into the same observable for site $k+1$ (mod $L$) one has
\be 
\label{C2}
\mbox{\bf C2:} \hspace*{10mm} T \rho T^{-1} = \rho, \quad T H T^{-1} = H. \hspace*{10mm} 
\ee 
Then $B^{\alpha\beta}_L(r,t) = C^{\alpha\beta}_L(0,r,t)$ and
$\tilde{B}^{\alpha\beta}_L(r,t) = \tilde{C}^{\alpha\beta}_L(0,r,t)$ 
and \eref{ccce3} becomes
\be
\exval{j^\alpha_k(t) q^\beta_{l+1}(0)}^c_L - \exval{q^\alpha_{k}(t) j^\beta_{l}(0)}^c_L 
= \frac{1}{L} \alpha^{\alpha\beta}_L
\ee
with the constant $\alpha^{\alpha\beta}_L$ given in \eref{alpha}.

We note that condition C2 together with C1 (decay of correlations) yields
\be 
\exval{j^\alpha_{k}(t) q^\beta_{0}(0)}^c_\infty = 
\exval{q^\alpha_{k+1}(t) j^\beta_{0}(0)}^c_\infty 
\ee
which is the lattice analogue of the local charge-current correlation equality
derived for translation invariant systems in continuous space in \cite{xxx}.

\subsubsection{Mutually commuting charges}

We finally comment on mutually commuting charges where
\be 
\label{C3}
\mbox{\bf C3:} \hspace*{10mm} \comm{Q^\alpha}{Q^\beta} = \comm{Q^{\alpha}}{\rho} = 0 \hspace*{10mm} 
\ee 
for a set of charges labelled by $\alpha,\beta$.

(1) First we consider a canonical ensemble where the density matrix is build from eigenstates
of the conserved charges $Q^\alpha$ and $Q^\beta$, i.e., 
$Q^{\alpha,\beta} \rho = \rho Q^{\alpha,\beta} = L q^{\alpha,\beta} \rho$
with the charge densities $q^{\alpha,\beta}$. Then $a^{\alpha\beta}_L = \tilde{a}^{\alpha\beta}_L=0$
and \eref{ccce3} yields
\bel{ccce31}
B^{\alpha\beta}_L(r+1,t) =  \tilde{B}^{\alpha\beta}_L(r,t) 
\ee
for all $r$ and $t$ and any finite $L$, without assuming decay of correlations or
translation invariance.

(2) Second, we consider a generalized Gibbs ensemble of the form
\bel{GGE}
\tilde{\rho} = \frac{1}{Z} \rho \rme^{\sum_{\alpha=1}^n \lambda_\alpha Q^\alpha}
\ee
with $Z = \tr (\rho \rme^{\sum_{\alpha=1}^n \lambda_\alpha Q^\alpha})$ and 
stationary $\rho$ independent of the generalized chemical potentials $\lambda_\alpha$. 
It has been conjectured that such a 
GGE state emerges asymptotically in time  when an integrable system, which has 
an extensive number of conserved local charges, has suffered a sudden quench 
\cite{Rigol06,Rigol07}, see \cite{Polkov11} for a general review.
This conjecture has been checked explicitly in many non-interacting models, 
see for example \cite{Cala11,Collura14}, and tested in truly interacting 
integrable models with a truncated GGE taking into account only a finite number 
$n$ of charges \cite{Fagotti13,Fagotti14,Pozsgay13,rem}. Indirect experimental
evidence was found by Vidmar et al. \cite{Vidm15} who confirmed the 
magnetization profile that was theoretically predicted for the evolution
of the $XX$ quantum chain after a quench to a step initial state \cite{Anta99}
see also \cite{Mori19} on the current fluctuations in this setting. 

Given a GGE satisfying C3, which by construction (S1 and S2) is then also stationary, 
one has with the short-hand notation $\partial_\alpha \equiv \partial / (\partial \lambda_\alpha)$
\be
\partial_\alpha \ln{Z} = \exval{Q^\alpha}, \quad \partial_\alpha \exval{O} = \exval{OQ^\alpha}^c.
\ee
Thus one can express the constants $a^{\alpha\beta}_L$ and
$\tilde{a}^{\alpha\beta}_L$ as derivatives as
\be 
a^{\alpha\beta}_L = \partial_\beta j^\alpha, \quad \tilde{a}^{\alpha\beta}_L = \partial_\alpha j^\beta
\ee
and obtains from \eref{alpha2}
\bel{ccce31b}
\partial_\beta j^\alpha - \partial_\alpha j^\beta = L[B^{\alpha\beta}_L(\floor{L/2}+1,0) 
-  \tilde{B}^{\alpha\beta}_L(\floor{L/2},0)].
\ee

We note that condition C3 for the GGE together with the condition C1' \eref{C1s} 
on the decay of correlations yields the current symmetry
\bel{cursym}
\partial_\beta j^\alpha = \partial_\alpha j^\beta .
\ee
where the stationary expectations 
$j^\alpha$ are understood as functions of the generalized chemical
potentials $\lambda_\alpha$. No translation invariance is required.

The current symmetry \eref{cursym} appears in many contexts in hydrodynamic theory, 
see e.g. \cite{Spoh14,Spoh16} for a review and \cite{Cast16,Doyo17,xxx}
for recent applications in generalized hydrodynamics where it was derived 
under the assumption C1' (decay of correlations in the form \eref{C1s}), 
C2 (translation invariance) and C3 (GGE). A mathematically rigorous proof of this 
current symmetry in the classical Markovian context
was presented earlier in \cite{Gris11}, using the same conditions C1' and C2 
and arguments for the proof that were later employed in similar form
in \cite{Spoh14,Cast16}.

We also note that assumption C3 implies for the
grandcanonical ensemble the relation $a^{\alpha\beta}_L = \tilde{a}^{\beta\alpha}_L$ 
and hence for $\alpha=\beta$ the exact charge-current correlation equality \eref{ccce31}
for all $r$ and $t$ and any finite $L$.

\section{Linear response symmetries}
\label{Sec:LR}

We point out some straightforward consequences of the charge-current correlation
equalities for linear response in far-from-equilibrium systems. For definiteness, we 
assume conditions C1' (decay of correlations \eref{C1s}) and C2 (translation invariance)
to be satisfied.

Consider a time-dependent perturbation of the form
$H(t) = H_0 + h A(t)$
where $h$ is the interaction strength. The linear-response function 
for an observable $B$ is given by \cite{LRT}
\bel{defRF} 
\hat{R}_{AB}(t) := \frac{\rmd}{\rmd h} \exval{B(t)}\Bigr \rvert_{h=0} .
\ee
For a pulse at time $t_0=0$, 
i.e., when the perturbation is of the form
$A(t) = A \delta(t)$, 
and for a density matrix $\rho$ that is stationary under the evolution of
$H_0$, straightforward computation yields $\hat{R}_{AB}(t) = R_{AB}(t)\Theta(t)$
where \cite{LRT}
\be 
\label{Rstat}
R_{AB}(t) = \frac{i}{\hbar} \tr \left\{\rho \comm{A}{B(t)} \right\} 
\ee
with the time-dependent operator $B(t)$.

Consider now the response at site $k$ of the observable $B=q^\beta_k$ to a pulse 
perturbation with $A = q^\alpha_0$ at the origin. Then \eref{Rstat} yields
\be 
\label{Rabk}
R^{\alpha\beta}(k,t) = \frac{i}{\hbar} \tr \left\{\rho \comm{q^\alpha_0}{q^\beta_k(t)} \right\} .
\ee
The total response
\be 
R^{\alpha\beta}_0 :=  \sum_k R^{\alpha\beta}(k,t) 
= \frac{i}{\hbar} \tr \left\{\rho \comm{q^\alpha_0}{Q^\beta} \right\} = - R^{\beta\alpha}_0
\ee
is trivially antisymmetric in the indices and independent of time because of the 
conservation law. 

Now consider the first moment
\bea 
R^{\alpha\beta}_1(t) & := & \sum_{k=\floor{-L/2}+1}^{\floor{L/2}}  k R^{\alpha\beta}(k,t) 
\eea
which provides information about the position at time $t$ 
of the center of mass of the perturbation.
Taking the time-derivative and using decay of correlations \eref{C1s} yields a first moment
\be
v^{\alpha\beta} := \dot{R}^{\alpha\beta}_1(t) = \exval{\comm{Q^\alpha}{j^\beta(0)}}_{c}
\ee
that does not depend on time so that $R^{\alpha\beta}_1(t) = R^{\alpha\beta}_1(0)
+ v^{\alpha\beta} t$ holds exactly.

Furthermore, from the global correlation 
equality \eref{CEglob} one derives the symmetry property 
\bel{vab}  
v^{\alpha\beta} = - v^{\beta\alpha}
\ee
between the first moments. Assuming further
condition C3 (commutativity of the conserved charges with the stationary density matrix),
one obtains 
\bel{vab2}
v^{\alpha\beta} = v^{\beta\alpha} = 0, 
\ee
both in the canonical and grandcanonical ensemble.

\section{Conclusions} 

The charge-current correlation equalities \eref{ccce1}, \eref{ccce2}, \eref{ccce2t}, and 
\eref{ccce3}  - \eref{alpha2} are generally valid, without any specific hypothesis on the nature of
a stationary quantum system with conserved charges $q^\alpha_k(t)$ that satisfy the 
discrete continuity equation \eref{S2} and have finite stationary cross correlations
among themselves and with the currents $j^\alpha_l(t')$. More specialized equalities
arise as when conditions C1 \eref{C1} or C1' \eref{C1s} 
on the decay of correlations are assumed to hold (see \eref{ccce32}, \eref{ccce33}, and \eref{vab}) 
or if some of the conserved charges commute among themselves 
and with the stationary density matrix (see \eref{ccce31}, \eref{ccce31b}, the current
symmetry \eref{cursym}, and the linear response symmetry \eref{vab2}). Translation 
invariance does not play a role for the validity of these correlation equalities.

These results clarify and generalize the range of validity of similar relations 
obtained in \cite{Cast16,xxx} for translation invariant systems. 
The correlation equalities are valid arbitrarily far from thermal equilibrium and 
provide concrete information about the 
spatial structure of the linear response function under these general conditions and about
finite-size corrections involving the local charge-current correlations.

As pointed out in \cite{Toth03}, the current symmetry \eref{cursym} guarantees that 
for stationary GGE's only hyperbolic systems of conservation laws can arise as 
hydrodynamic limits that govern the macroscopic time-evolution of the local conserved 
quantities. When the fluctuations of the locally conserved charges are the most relevant 
slow dynamical variables, one expects in one space dimension from mode-coupling theory 
\cite{Spoh14} that fluctuations around the deterministic hydrodynamics
are generically diffusive or in the Kardar-Parisi-Zhang (KPZ) universality class \cite{Halp15}
and, on special manifolds in the space of densities and model parameters,
in the Fibonacci universality classes \cite{Popk15b} which include the diffusive
and superdiffusive Kardar-Parisi-Zhang universality class as paradigmatic members.
For recent evidence of diffusive and superdiffusive transport we mention 
\cite{Ilie18,Gopa19} and more specifically on the observation of KPZ physics
in the SU(2) symmetric Heisenberg spin chain we refer to \cite{Ljub19,DeNa19}.

Finally, we note that the current symmetry \eref{cursym} may be useful in 
numerical computations of quantum quenches as a probe of an underlying asymptotic 
GGE, as \eref{cursym} would not be
valid if the local stationary state does not approximate a GGE. Likewise, the
linear response symmetry \eref{vab2} can be used as a probe of the symmetries of a
density matrix when its only {\it a priori} known property is stationarity.

It is a pleasure to acknowledge stimulating discussions with
A. Kl\"umper and V. Popkov and to thank B. Doyon for helpful comments on an 
earlier version of this paper. This work was
supported by Deutsche Forschungsgemeinschaft.
G.M.S. thanks the Laboratoire de Physique et Chimie Th\'eoriques,
Universit\'e de Lorraine, where part of this work was done, for kind hospitality.

\end{document}